# A New Approach toward Transition State Spectroscopy

**Kirill Prozument**[a1], **Rachel Glyn Shaver**[a2], **Monika A. Ciuba**[b], **John S. Muenter**[c], **G. Barratt Park**[a3], **John F. Stanton**[d], **Hua Guo**[e], **Bryan M. Wong**[f], **David S. Perry**[g], and **Robert W. Field**[a4*]

*a) MIT, Department of Chemistry, Cambridge, MA 02139 1) kuyanov@mit.edu, 2) rshaver@mit.edu, 3) barratt@mit.edu, 4) rwfield@mit.edu, b) Arizona State University, Department of Chemistry and Biochemistry, Tempe, AZ 85287-5601 monika.ciuba@asu.edu, c) University of Rochester, Department of Chemistry, Rochester, NY 14627 muenter@chem.rochester.edu, d) University of Texas at Austin, Department of Chemistry and Biochemistry, Austin, TX 78712 jfstanton@mail.utexas.edu, e) University of New Mexico, Department of Chemistry, Albuquerque, NM 87131 hguo@unm.edu, f) Sandia National Laboratories, Materials Chemistry Department, Livermore, CA 94550 usagi@alum.mit.edu, g) The University of Akron, Department of Chemistry, Akron, OH 44325-3601 dperry@uakron.edu*, * corresponding author.

Chirped-Pulse millimetre-Wave (CPmmW) rotational spectroscopy provides a new class of information about photolysis transition state(s). Measured intensities in rotational spectra determine species-isomer-vibrational populations, provided that rotational populations can be thermalized. The formation and detection of $S_0$ vinylidene is discussed in the limits of low and high initial rotational excitation. CPmmW spectra of 193 nm photolysis of Vinyl Cyanide (Acrylonitrile) contain J=0-1 transitions in more than 20 vibrational levels of HCN, HNC, but no transitions in vinylidene or highly excited local-bender vibrational levels of acetylene. Reasons for the non-observation of the vinylidene co-product of HCN are discussed.

## A. Introduction

Chirped-Pulse millimetre-Wave (CPmmW) spectroscopy[1-3] is capable of determining the relative species-conformer-vibrational level populations of all polar products of a photolysis reaction. These experimentally determined populations encode the structures of the transition states that are most important at each photolysis wavelength or combination of wavelengths.[4] The isomer-conformer-vibrational level population information obtained from a CPmmW spectrum is more complete than what is obtainable by mass spectrometry[5] and free of the need for the transition strength and quantum yield determinations that are required for most laser-based population measurements. However, difficulties exist in the use of populations determined by CPmmW spectroscopy to characterize transition states.

This paper is a case study of the application of CPmmW spectroscopy to the 193 nm photolysis of vinyl cyanide (acrylonitrile): $CH_2=CHCN$.[5] The most important results are the observation of many vibrationally excited levels of HCN[6,7] and HNC[7,8], the comparison of the observed isomer and vibrational population distribution when $CH_2=CHCN$ is replaced by $CH_2=CDCN$, the dependence of the population distribution on the amount of post-photolysis rotational cooling, and the failure to observe either $S_0$ vinylidene or extreme local-bender vibrationally excited $S_0$ acetylene. Both vinylidene and local-bender acetylene have large electric dipole moments along the inertial a-axis, which are expected to give rise to strong J=0-1, $K_a$=0-0 transitions in the 70-90 GHz region of our CPmmW spectrometer.

## B. Chirped Pulse Spectroscopy

Chirped Pulse Fourier Transform Microwave Spectroscopy is a revolutionary technique developed in the research group of Brooks Pate at the University of Virginia.[2,3] In its initial implementation, the frequency of a microwave pulse is chirped linearly in time over several GHz. This microwave pulse is broadcast into a gas phase molecular sample, polarizing all two-level systems with frequencies within the spectral interval of the chirped pulse. These polarizations relax by Free Induction Decay (FID), which is a voltage vs. time signal that is collected, down-converted by mixing with a local oscillator (heterodyne detection), and recorded in the time-domain in a fast oscilloscope. All components of the experiment are synchronized to a 10 MHz master clock, thus the FIDs from all chirps are perfectly phase coherent and may be averaged in oscilloscope memory. The detected signal scales as[9]

$$P_{1,2} \propto \frac{\mu^2 \varepsilon_0^2 N_{1,2}}{\alpha^{1/2}} A^{1/2} L \qquad (1)$$

where $\mu$ is the electric dipole moment, $\varepsilon_0$ is the peak microwave electric field, $\Delta N_{1,2}$ is the population density difference between levels 1 and 2, $\alpha$ is the chirp rate, $A$ is the cross-sectional area of the microwave beam, and $L$ is the length of the irradiated region. A chirp over 10 GHz can be completed in a time between 10 ns and 10 μs. The advantage of a slow chirp (large $\alpha^{-1/2}$) is limited by the necessity that the CP terminate before the FID has decayed significantly. The FID is collected over a time interval from ~1 μs to 20 μs, where the decay rate of the FID is determined by inelastic collisions ($T_1$, typically 10 MHz/Torr) or Doppler dephasing ($T_2 \propto \Delta E_{1,2}^{-1} T^{-1/2}$). Usually, CP spectra are recorded at sufficiently low pressure that $T_1$ is slightly longer than $T_2$.

When the time-domain signal is transformed into the frequency-domain, linewidths in the 100 to 500 kHz range are routinely obtained. A single chirp can contain ~$10^5$ resolution elements.[1-3] Since all two-level systems within the chirp frequency region are sampled in each chirp, relative transition intensities may be measured with accuracies approaching the 1% level, even for transition frequencies separated by 10 GHz.

The microwave electric field, $\varepsilon_0$, scales as $I^{1/2}$, where $I$ is the intensity in units of power/area. If the diameter of the microwave beam is increased by a factor of 10, $\varepsilon_0$ decreases by a factor of 10 but $\Delta N_{1,2} A^{1/2}$ increases by an exactly cancelling factor of 10. Therefore there is no penalty in choosing the cross-section of the microwave beam to be as large as that of the molecular sample. However, optimum CP signals are obtained when the active volume is as long as possible. In the $\Delta E_{1,2} < k_B T$ limit, the Boltzmann contribution to the CP signal scales as $\Delta E_{1,2}/T^2$. For pure rotational CP spectroscopy of photolysis products, rotational thermalization to $T \approx$ 5K is very important.

In NMR there is a simple relationship between the integrated intensity of a single resolved transition and the number of nuclei in the chemical environment that corresponds to the selected chemical shift.[10] In microwave spectroscopy there is a similarly simple relationship between the transition intensity and the population difference between the upper and lower levels connected by a selected transition frequency.[11] One needs to know the rotational temperature, the permanent electric dipole moment, and the transition width. These three quantities are routinely measured by microwave spectroscopy. However, two fundamental problems complicate the use of CP spectroscopy to measure populations of very small molecules, especially when the goal is to determine the nascent, non-thermalized, species-conformer-vibrational level populations of reaction products. (i) The rotational constants of small (near prolate top, a-dipole) molecules are typically so large that *only* the lowest J↔J+1 transition lies within the operation range of the spectrometer. Thus, unless one assumes that rotation of the target molecule is thermalized at a temperature determined by exploiting the measured rotational population distribution of a heavier "thermometer molecule" or from the translational temperature determined from a Doppler linewidth, one cannot derive vibrational populations from a single measured rotational population difference. (ii) Since rotation-thermalizing transitions occur typically at a rate $10^3$ times that of vibration-thermalizing transitions[12], it is plausible that in a supersonic jet expansion, rotational populations can be thermalized yet vibrational populations will be near-nascent. However, when an enormous amount of reaction exoergicity is deposited in rotation, it is dangerous to make any kind of assumption about rotational thermalization.

**C. $S_0$ Vinylidene: C=CH$_2$**

On the acetylene electronic ground state potential energy surface ($S_0$), linear acetylene is separated by an ~ 2 eV barrier from the $C_{2v}$ symmetry vinylidene isomer[13,14]. Although vinylidene is a widely invoked reaction intermediate, vinylidene has never been chemically isolated. There have been definitive spectroscopic observations of $S_0$ vinylidene[15-17], but these observations have often been misinterpreted. From the vinylidene side, the barrier to isomerization to acetylene is very low, ≲ 0.09 eV. However, the acetylene↔vinylidene saddle point lies well below the lowest dissociation limit at 5.6 eV[18], thus all of the information about isomerization rates and mechanisms is encoded in energy-resolved eigenstates[16], in particular in the distribution of acetylene and vinylidene basis state characters in each eigenstate. It is *not* a textbook style bound↔free barrier-tunnelling problem.

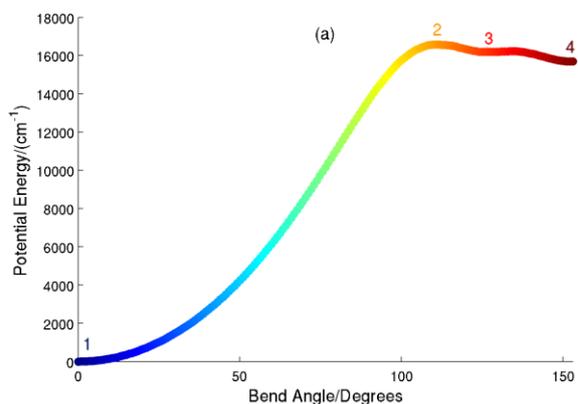

Figure 1. The Minimum Energy Isomerization Path between Acetylene and Vinylidene. The one-dimensional path is colour-coded to correspond to the dependence of the molecular geometry, shown in Fig. 2, on the progress along the isomerization path, which is predominantly a local CCH bend. The numbers 1-4 correspond to the locations of the two H atoms along the isomerization path, the structure of which is also labelled on Fig. 2.[19]

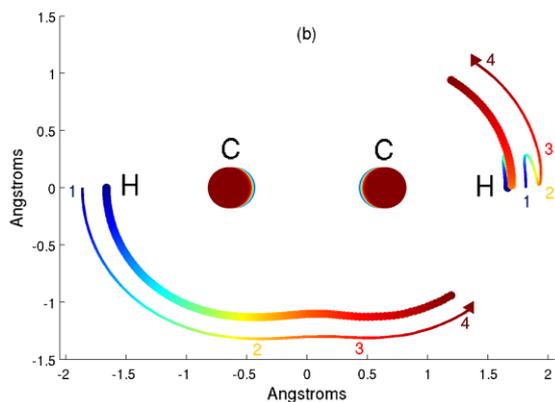

Figure 2. Geometric structure of $S_0$ HCCH as a function of the (colour-coded) progress along the isomerization path shown in Fig. 1[19].

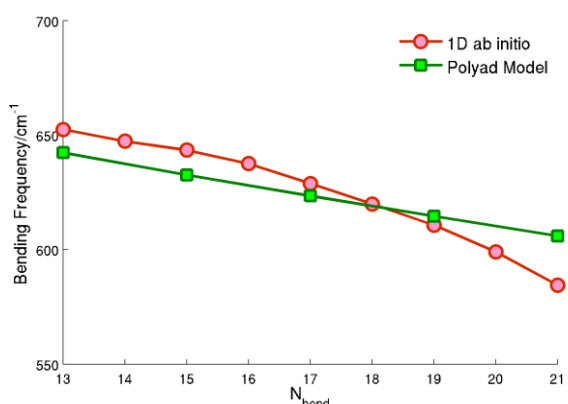

**Figure 3:** Comparison of the Frequency of the *ab initio* One-Dimensional Local-Bending Vibration with the Frequency Obtained from the Low-J $\mathbf{H}^{\text{eff}}$ Fit Model[21]. The agreement is satisfactory in the region between the emergence of the local-bend and $N_{bend} \approx 14$ at the top of the isomerization barrier near $N_{bend} \approx 24$[19,20].

As shown in Figs. 1 and 2, the minimum energy isomerization path from linear acetylene to vinylidene is almost exclusively a large amplitude acetylene local-bend (i.e. one CCH bend, not one of the two bending normal modes, $\nu_4$-trans bend or $\nu_5$-cis bend)[19,20]. This suggests that, in the absence of strong rotation-induced "Coriolis" or "rotational $\ell$-resonance" mixing[21-24] [two different forms of rotationally mediated interactions, referred to hereafter as Coriolis] of highly excited acetylene basis states, one expects that vibrationally unexcited vinylidene will interact exclusively with the very sparse manifold of highly excited local-bender levels of acetylene. Local-bender levels correspond to roughly 1% of the total vibrational density of states at the vibrational excitation energy of the barrier. Thus nonrotating $C_2H_2$ will encode isomerization as few-level vinylidene↔local-bender acetylene quantum beats, but highly rotating $C_2H_2$ might present to each vinylidene basis state the full (symmetry-sorted) density of acetylene eigenstates (about one vibrational level per 7 cm$^{-1}$), each with a small fractional local-bender character. Thus, profoundly different limiting behaviours might be expected, dependent on whether the energized species is formed rotationally hot or rotationally cold.

There are two classes of experiments that seem to have sampled these two limits. The rotationally cold limit is realized by **photodetachment** of the extra electron from the vinylidene negative ion.[15-17] $C=CH_2^-$ has geometric and vibrational structure almost identical to that of $S_0$ vinylidene. Ejection of a light electron cannot be accompanied by significant rotational excitation of the heavy molecule. In contrast, formation of vinylidene by **photolysis** of vinyl cyanide[5], vinyl bromide, vinyl chloride[25], or propene creates a highly rotationally excited $C_2H_2$ fragment. Even if, at high-J, the lowest energy vibrational levels of vinylidene are not affected by Coriolis mixing, the highly excited vibrational levels of acetylene formed by photolysis of vinyl-X are profoundly Coriolis-mixed. The acetylene local-bender basis state character, which serves to vinylidene as the unique doorway into acetylene, is admixed at high-J into the entirety of the dense manifold of vibrationally excited acetylene levels. This rotation-induced mixing (exclusively on the acetylene side of the barrier) serves as the on-switch for the more familiar mechanism of isomerization into a dense and ergodic manifold of vibrational states.

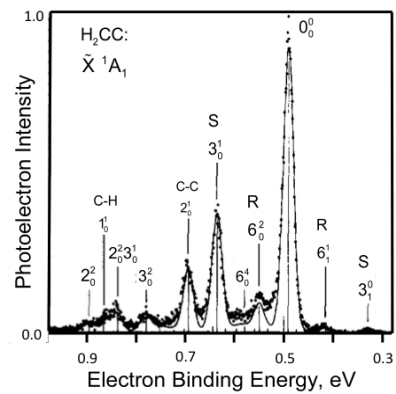

**Figure 4:** The Photoelectron Spectrum of the Vinylidene Negative Ion. The widths of each vibrational band feature are only slightly larger than the computed unresolved rotational structure at $T_{rot}$ equal to $T_{trans}$ of the Helium carrier gas, upper right panel of Fig. 5 of Ervin et al.[16]

The spectrum and dynamics of $S_0$ vinylidene became a subject of great excitement and controversy when the Lineberger research group observed and assigned the lowest few vibrational levels of $S_0$ vinylidene in the photoelectron spectrum (PES) of the $H_2C=C^-$ negative ion[15-16]. A fixed frequency UV line of an Argon ion laser was used to detach the extra electron from $H_2C=C^-$ and the difference between the energy of the UV photon and the kinetic energy of the detached electron provided information about the energies and widths of the $S_0$ vinylidene vibrational "resonances." The resolution of the PES was insufficient to resolve the J, $K_a$ rotational structure of these vibrational resonances in this near-prolate asymmetric top. In the original paper, *most* of each observed linewidth was explicitly and correctly interpreted as inhomogeneous broadening due to unresolved rotational structure[16]. Despite this carefully crafted explanation, many papers from other research groups were quick to interpret *all* of the resonance width to homogeneous broadening due to tunnelling through a low barrier into either a continuum or vibrational quasi-continuum of $S_0$ acetylene. Lifetimes of $S_0$ vinylidene vibrational levels ranging from a few picoseconds to as short as ~100 femtoseconds have been cited as having been "experimentally measured" in the PES spectrum[16]. There is a third possible explanation, strongly supported by Coulomb explosion experiments[17], that the excess width is due to few-level mixing between one zero-order vinylidene vibrational basis state and one or two near-degenerate acetylene vibrational basis states that have some extreme local-bender character.

Like the Lineberger group negative ion PES spectra[15,16], the Coulomb explosion experiments[17] start with the $H_2C=C^-$ negative ion. This negative ion is accelerated to very high kinetic energy and impacted onto a thin foil. This impact results in the removal of all electrons. After impact (the Coulomb explosion) the positions and arrival times of the resultant two $H^+$ and two $C^{6+}$ cations (*all four ions* from *each* parent molecule) are recorded, event by event, on a multi-particle detector. From the arrival times and positions, the structure of the parent molecule is back-calculated at the instant of collision with the foil. The extraordinary thing about the Coulomb explosion experiments is that, even when the extra electron on the $H_2C=C^-$ negative ion is detached 3.5 μs before the parent molecule collides with the foil, 50% of the reconstructed images are of the vinylidene structure[17]. One explanation that accounts for the extra width in the negative ion PES and the long-time persistence of the vinylidene structure is a few-level quantum beating system where the long-lived, eigenstates are mixtures of one vinylidene and one or two acetylene basis states with admixed local-bender character. Interestingly, the *initial* decay of vinylidene character in this quantum-beating system could be on the subpicosecond time-scale, but the dynamics must be periodic rather than irreversible.

Detachment of an electron from the $H_2C=C^-$ negative ion results in much less vibrational and rotational excitation than ultraviolet photolysis of vinyl-X (X = Cl, Br, CN, $CH_3$). For photolysis, this is partly a kinematic effect and partly a Franck-Condon mediated energetic effect. There is an enormous difference in nuclear geometry between the $S_0$ zero-point vibrational level of the parent molecule and the geometry of the conical intersection that feeds the photofragmentation transition states. This ensures, due to Franck-Condon restrictions, that UV photolysis occurs at energies far above the thermodynamic threshold. The photofragments are formed significantly rotationally and vibrationally excited. This is bad news for high resolution spectroscopic measures of energy partitioning, because population is distributed over many rotation-vibration levels, individual rovibronically resolved transitions will be weak, and the spectrum will be complex and congested. Time-resolved FTIR spectroscopy is well suited to the characterization of photolysis product distributions[26]. The relatively low resolution of TR-FTIR makes it very difficult to determine rotational population distributions. However, very strong Δv=-1 vibrational propensity rules for infrared active normal modes often usefully simplify the low-resolution vibrational spectrum. For example, in $S_0$ acetylene, only $\nu_3$ (antisymmetric CH stretch, parallel type) and $\nu_5$ (*cis*-bend, perpendicular type) are infrared active and have recognizably different rotational band contours.

The Hai-Lung Dai group has applied Time-Resolved-FTIR spectroscopy to the 193 nm photolysis of vinyl-X molecules[26]. They observe spectrally distinct vibrational features that enable measurement of the HCN:HNC product ratio. Surprisingly, one of the strongest vibrational features in their early-time spectrum cannot be assigned as one of the known IR active fundamentals of acetylene, vinylidene, HCN, or HNC. They assign it as the acetylene $\nu_4+\nu_5$ combination band (perpendicular type) and attribute its presence in the spectrum as due to collision-induced decay of vinylidene into high vibrational levels of acetylene[5,26,27]. Combination bands are usually weak relative to IR active fundamentals, even when the combination band originates from a high-$(v_4,v_5)$ level. These features are more likely assignable as Δv=-2, Δ$\ell$=0 local-bender overtone bands (parallel type), which arise directly from the decay of vinylidene along the isomerization path[28] into highly excited acetylene local-bender levels. The choice between these competing assignments might eventually be settled by observation of the rotational band contour, which is very different for parallel (no Q branch) vs. perpendicular (strong Q branch) type bands.

The low-resolution of the TR-FTIR spectra will preserve the intensity associated with emission from highly excited local-bender levels, even in the presence of strong Coriolis mixing of the local-bender character into a quasi-continuum of highly excited vibrational levels. The mixing does not cause the local-bender character to vanish. It dilutes it into many vibrational levels. The local-bender character will light up low-resolution spectra, but it will be diluted into unobservability in an eigenstate-resolved, high-resolution spectrum.

**D. Photolysis of Vinyl Cyanide. Mass Spectrometry and Photofragment Translational Spectroscopy**.

When vinyl cyanide is photolyzed by 193 nm radiation, the following photofragmentation products are energetically accessible[5]:

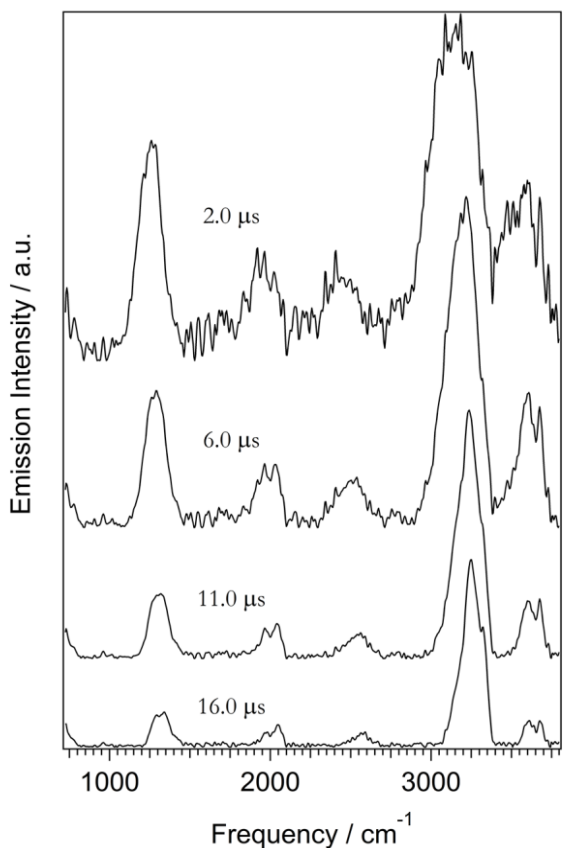

**Figure 5**: The Time-Resolved Fourier Transform Infrared Fluorescence Spectrum (TR-FTIR) Resulting from 193 nm Photolysis of Vinyl Cyanide. The strong early-time feature near 1350 cm$^{-1}$ is assigned here as superimposed acetylene $\Delta v=-2$ local-bender (parallel type) transitions and in the original publication as $\nu_4+\nu_5$ (perpendicular type) combination bands. This is Figure 1 from Wilhelm et al.[26]

| | | |
|---|---|---|
| $H_2C=C-C\equiv N$ | (m/e 52) + H | (m/e 1) |
| $HC=CHC\equiv N$ | (m/e 52) + H | (m/e 1) |
| $H_2C=CH$ | (m/e 27) + $C\equiv N$ | (m/e 26) |
| $HC=C-C\equiv N$ | (m/e 51) + $H_2$ | (m/e 2) |
| $^1(C=CHC\equiv N)$ | (m/e 51) + $H_2$ | (m/e 2) |
| $^3(C=CHC\equiv N)$ | (m/e 51) + $H_2$ | (m/e 2) |
| $HC\equiv CH$ | (m/e 26) + HCN or HNC | (m/e 27) |
| $^1(H_2C=C)$ | (m/e 26) + HCN or HNC | (m/e 27) |
| $^3(H_2C=C)$ | (m/e 26) + HCN or HNC | (m/e 27) |

Provided that vacuum ultraviolet radiation at sufficiently short wavelength is available, each of these photofragments may be ionized, mass selected, and detected. The time-of-flight distribution from photodissociation to ionization provides the translational energy distribution, P(E$_{Trans}$), for each mass-selected photofragment. The P(E$_{Trans}$) contains information about the nascent vibrational energy distribution (unaffected by post-photolysis collisions) within each photofragment. Mass spectrometry alone is not sufficient to distinguish two different chemical species (e.g. H$_2$C=CH vs. HCN or HC≡CH vs. C≡N) [except at extremely high mass resolution, which is starting to be available at national user facilities] or different isomers (e.g. HCN vs. HNC or HC≡CH vs. H$_2$C=C) with the same m/e. However, different chemical species can usually be distinguished via their distinct ionization thresholds[5]. Whether isomers can be distinguished depends on whether the unstable isomer decays on the time-scale of the photofragmentation half-collision. HNC is sufficiently stable that its contribution to the m/e 27 P(E$_{Trans}$) can be distinguished from that of HCN. However, the isomerization rate of vinylidene has been a subject of controversy and also the motivating factor behind the long-term effort in the Field group to spectroscopically characterize vinylidene.

There is an enormous difference in the amount of rotational excitation of vinylidene formed by photodetachment of an electron from the vinylidene negative ion vs. vinylidene formed by photolysis of vinyl cyanide. Rotation matters! As demonstrated by the **H**$^{eff}$ of Perry et al.[24], the exceptional stability of low-J acetylene extreme local-bender states[21-23] is destroyed at high J (see Figs. 9 and 10 in Section E). At high-J, the local-benders undergo rapid Intramolecular Vibrational Redistribution into the dense and ergodic manifold of highly excited vibrational levels. The minimum energy isomerization path between acetylene and vinylidene lies along the local-bend from the acetylene side[19] and the in-plane wag from the vinylidene side[16] of the isomerization barrier. Nonrotating vinylidene interacts exclusively with isoenergetic local-bender levels of acetylene. The density of acetylene local-bender vibrational states is very low, typical of a *diatomic* molecule. The Coulomb explosion experiments (that also form low-J vinylidene by photodetachment of an electron from the vinylidene negative ion[17]) show that the acetylene-vinylidene interaction is that of few-level quantum beats. At low-J, there is no irreversible isomerization from vinylidene into a dense manifold of highly excited vibrational levels of acetylene.

The negative ion PES experiments in the Lineberger group[16] reveal vinylidene vibrational resonances for which most of the width is due to rotational inhomogeneous broadening. The extra width could well be due to unresolved splittings between pairs of locally-perturbing levels rather than a Fermi Golden Rule broadening due to mixing into a vibrational quasi-continuum. The fast decay of vinylidene character could well be the initial (reversible) decay of a two-level quantum beating system. Yet Ervin et al.[16] write "The width of the singlet transition yields a lower limit to the vinylidene lifetime τ>0.027 ps. From an analysis of line shapes including instrumental and rotational broadening, we estimate the lifetime of the singlet to be τ~0.04-0.2 ps." This corresponds to FWHM of 250-50 cm$^{-1}$. [Earlier in their paper, Ervin et al.[16] presciently state "Since the acetylene--vinylidene system is bound, in a much higher resolution experiment there would be no 'lifetime broadening' as such. Rather, the zero-order vibrational levels we identify with vinylidene would be split into many eigenlevels of the complete system."] This lifetime estimate is echoed by Blank et al.[5], "Based on negative ion photodetachment spectral linewidths, Lineberger and co-workers estimated a lifetime of 40-200 fs for the isomerization of vinylidene to acetylene."

Everyone is correct. At low-J, as in photodetachment experiments, vinylidene lives longer than 3.5 μs. Its lifetime is probably on the same time-scale as a typical acetylene local-bender $v_{bend}$=24-22 vibrational radiative decay, ~1 ms. At high-J, as in photolysis experiments, the vinylidene lifetime is acetylene IVR limited. 40-200 fs is probably an excellent estimate.

Zou et al.[13] report full 6-dimensional calculations of the J=0 vibrational levels in the energy region from ~500 cm$^{-1}$ below the vinylidene zero-point level to ~1800 cm$^{-1}$ above the acetylene-vinylidene saddle point. They find *isolated* vibrational eigenstates that have predominant vinylidene character (shown in their Figs. 2, 3, and 6) at an ~1 per 7 cm$^{-1}$ symmetry-sorted ($A_1'$ in the Complete Nuclear Permutation Inversion Group, $G_8$) vibrational density of states. The total vibrational density of states is four times larger than 1/7cm$^{-1}$ because only four of the eight CNPI symmetry species have a nonzero nuclear statistical weight. Since the tunneling interactions between acetylene and vinylidene vibrational basis states are diagonal within each CNPI symmetry species, 1/7cm$^{-1}$ is the density of states that is relevant to a Fermi Golden Rule estimate of the interaction matrix elements between acetylene and vinylidene basis states.

The FWHM associated with a state that has 0.2 ps lifetime is 50 cm$^{-1}$. There are only ~50/7 $A_1'$ vibrational eigenstates within this FWHM. Although each of these ~7 eigenstates is long-lived (longer than 3.5 μs), together they encode the decay of a single vinylidene basis state formed by the photodetachment event at t=0. Using the Fermi Golden Rule in reverse to estimate the average tunneling matrix element between one vinylidene basis state and each of the nearby acetylene basis states,

$$w = (2\pi/\hbar)\langle\phi_{vinylidene}|H|\phi_{acetylene}\rangle^2 \rho, \quad (2)$$

where $w$=1/0.2ps is the tunneling rate, $H$ is the interaction matrix element, and $\rho$=□□7cm$^{-1}$ is the vibrational density of states, then the average matrix element, in cm$^{-1}$ units, is 4 cm$^{-1}$. With such a small average tunneling matrix element, significant vinylidene~acetylene mixing (approaching 50:50) only occurs between nearest neighbor eigenstates. This is quite consistent with the Coulomb explosion result[17], in which images are observed consisting typically of 50% acetylene and 50% vinylidene structures.

**E. Spectroscopy of Acetylene**.

The vibration-rotation energy levels of the acetylene $S_0$ electronic state have been exhaustively examined in high-resolution (~0.001 cm$^{-1}$) infrared and Raman[29], moderate resolution (~0.03 cm$^{-1}$) $S_1$-$S_0$ Stimulated Emission Pumping (SEP)[30,31], and low-resolution (~5 cm$^{-1}$) $S_1$-$S_0$ Dispersed Fluorescence spectra[30-33].

As for any linear four-atom molecule, there are seven vibrational degrees of freedom, three nondegenerate stretching modes ($v_1$ symmetric CH stretch, $v_2$ CC stretch, $v_3$ antisymmetric CH stretch) and two doubly-degenerate bends ($v_4$ *trans*-bend and $v_5$ *cis*-bend). All of the normal mode vibrational frequencies are in approximate integer multiple ratios [($v_1$: $v_2$: $v_3$:$v_4$: $v_5$) = (5:3:5:1:1)][34,35]. Whenever vibrational frequencies are in near integer multiple ratios, second-order perturbation theory breaks down and quasi-degenerate perturbation theory requires diagonalization of an approximately block diagonal effective Hamiltonian matrix, $\mathbf{H}^{eff}$ [34]. Each block of this $\mathbf{H}^{eff}$ is called a **polyad**. The vast majority of vibrational levels are members of a polyad, and all of the normal mode vibrational quantum numbers ($v_1,v_2,v_3,v_4,\ell_4,v_5,\ell_5$) are destroyed by anharmonic interactions between quasi-degenerate basis states. The result of these inter-mode anharmonic interactions in $S_0$ acetylene is that the only surviving approximately good vibrational quantum numbers are the polyad quantum numbers: $N_{resonance}=5v_1+3v_2+5v_3+v_4+v_5$, $N_{stretch}=v_1+v_2+v_3$, and $\ell_{total}=\ell_4+\ell_5$ [34].

The total vibrational Hamiltonian is approximately block-diagonal and each block is a polyad that is denoted by its three polyad quantum numbers [$N_{res},N_s,\ell_{tot}$]. The intrapolyad matrix elements follow standard harmonic oscillator matrix element selection and scaling rules[34]. For example, the intrapolyad anharmonic matrix element $k_{1133}\mathbf{Q}_1^2\mathbf{Q}_3^2$ has intra-polyad selection rules $\Delta v_1=-\Delta v_3=\pm 2$ and 0 and the values of such matrix elements follow the scaling rule

$$\langle v_1 v_3|k_{1133}\mathbf{Q}_1^2\mathbf{Q}_3^2|v_1 v_3\rangle = \frac{1}{2}K_{1133}v_1 v_3$$

$$\langle v_1 v_3|k_{1133}\mathbf{Q}_1^2\mathbf{Q}_3^2|v_1+2,v_3-2\rangle = \frac{1}{4}K_{1133}[(v_1+1)(v_1+1)(v_3)(v_3-1)]^{1/2} \quad (3)$$

where $k_{1133}$ is the force constant for quartic interaction between modes 1 and 3 and $K_{1133}$ is the $v_1,v_3$-independent scaling parameter. The $k_{1133}\mathbf{Q}_1^2\mathbf{Q}_3^2$ term also gives rise to inter-polyad interactions, for example

$$\langle v_1 v_3|k_{1133}\mathbf{Q}_1^2\mathbf{Q}_3^2|v_1+2,v_3+2\rangle = \frac{1}{4}K_{1133}[(v_1+2)(v_1+1)(v_3+2)(v_3+1)]^{1/2} \quad (4)$$

but the energy denominator, $2\omega_1+2\omega_2$, permits treatment of such interpolyad effects by second-order perturbation theory. This makes it possible to replace the exact Hamiltonian by a lower-dimension block-diagonal $\mathbf{H}^{eff}$, where the membership and matrix elements in each polyad block are explicitly related to those in all other blocks.

In the special case that $N_s=0$, then $N_{res}=N_{bend}=v_4+v_5$. The $[N_{bend}=10,\ell_{tot}=0]$ polyad, for example factors into four blocks, $[10,0,g,e]$ (12 members), $[10,0,g,f]$ (6 members), $[10,0,u,e]$ (9 members), and $[10,0,u,f]$ (9 members).

The $\mathbf{H}^{eff}$ is a powerful and convenient fit-model, because it requires far fewer adjustable parameters than a complete, full-dimensional Hamiltonian, $\mathbf{H}$. The $\mathbf{H}^{eff}$ provides accurate predictions of the energies of unobserved vibrational levels, the eigenvector for each eigenstate expressed as a linear combination of normal mode product basis states that are members of the $[N_{res},N_s,\ell_{tot}]$ polyad,

$$\left|[N_{res},N_s,l_{tot}],\text{energy rank } i\right\rangle = \sum_{\substack{\text{basis state} \\ j}} c_j^i \left|[N_{res},N_s,l_{tot}],j\right\rangle \quad (5)$$

predictions of transition propensity rules and relative intensities, for example when the k-th basis state is "bright" and all of the other basis states in the polyad are "dark,

$$I_{[N_{res},N_s,N_{tot}],i} = \left(c_k^i\right)^2 \quad (6)$$

and predictions of inter-isomer interaction matrix elements, for example when only the k-th polyad basis state (e.g. an extreme local-bender) has significant overlap with the other-isomer basis state,

$$\left\langle [N_{res},N_s,N_{tot}],i\middle|\mathbf{H}\middle|\text{other isomer}\right\rangle \propto c_k^i \quad (7).$$

In addition to the complete Hamiltonian, $\mathbf{H}$, there are two classes of $\mathbf{H}^{eff}$ models for the vibration-rotation levels of acetylene: (i) the low-J $\mathbf{H}^{eff}$ that is determined primarily by fits to low-resolution SEP and DF data mostly obtained from polyads with significant excitation in the bending modes[21-23], and (ii) the all-J $\mathbf{H}^{eff}$ that is determined by global fits to high resolution spectra, accurate to measurement uncertainty, of all observed rotation-vibration energy levels[24].

The complete $\mathbf{H}$, derived from a high-level *ab initio* potential energy surface, can be an excellent approximation to "the truth," but it is not organized in a way that directly identifies the almost-good quantum numbers and all of the relevant matrix element propensity and scaling rules. The low-J $\mathbf{H}^{eff}$ is parametrically parsimonious, compact, and it explicitly reveals the scalable and mechanistically explicit forms of all spectroscopic and dynamical parameters[21]. However, since the low-J $\mathbf{H}^{eff}$ is determined from low-J data, it does not include J-dependent interaction mechanisms (in particular the rotational $\ell$-resonance term, the matrix elements for which scale as $J(J+1)$) and cannot be expected to scale reliably to high-J levels. The all-J $\mathbf{H}^{eff}$ is expressed in terms of an enormous number of fit parameters and requires numerical experimentation to reveal the goodness of approximately conserved observables, in particular how well quantities that are well conserved at low-J continue to be conserved at high-J[24].

A staggering amount of information is contained in the set of eigenvectors of the complete or an effective Hamiltonian. Two ways of looking for the emergence or disappearance of a class of localization in state space are the "participation number"[36] and a form of correlation diagram known as a "spaghetti diagram"[21]. The participation number is a measure of how many basis states have a significant presence in a specified eigenstate. For the i-th eigenstate the participation number, $P_i$, is

$$P_i = \left[\sum_j \left(c_j^i\right)^4\right]^{-1} \quad (8)$$

where the mixing coefficients, $c_j^i$, are defined in Eq. (5).

A $P_i$ value near 1 means that the i-th eigenstate is localized in a single basis state, whereas a $P_i$ value near 10 means that 10 basis states contribute significantly to the i-th eigenstate. [For example, if there are 10 basis states that contribute equally to the i-th eigenstate, then each mixing coefficient is $10^{-1/2}$ and $P_i=10$.] A complementary quantity, the basis state "dilution factor," $D_j$, may be defined similarly,

$$D_j = \left[\sum_i \left(c_j^i\right)^4\right] . \quad (9)$$

When the j-th basis state is diluted effectively into 10 eigenstates, $D_j=1/10$. A fundamental objection is that the $P_i$ and $D_j$, as defined in Eqs. (8) and (9), depend on the choice of basis set. Starting from the complete normal mode basis set, it is possible to define a variety of physically relevant basis sets by applying a unitary transformation that diagonalizes $\mathbf{H}^{(0)}$ plus one or more of the anharmonic resonance terms in $\mathbf{H}^{(1)}$ [24]. One choice is to transform from a normal mode to a local-mode Hamiltonian[21]. Each choice of basis set is capable of conveying different classes of insight into the nature of the dynamics in various regions of state space[24,37].

Figure 6 is a correlation (spaghetti) diagram that displays the goodness of the basis state quantum numbers in different energy regions of a polyad and also for high- vs. low-energy polyads[21]. The polyads shown are "pure-bending" polyads. They are pure-bending because, when $N_{str}=0$, then $N_{res}=N_{bend}=v_4+v_5$.

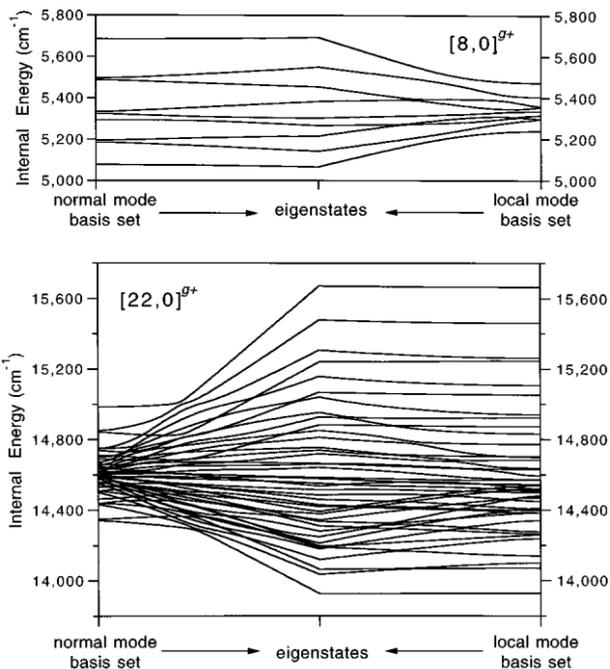

**Figure 6:** Correlation diagrams for the normal and local mode basis sets for the [$N_{bend}$=8,$\ell_{total}$=0, *gerade*, e-symmetry] (top) and [22,0,*g*,e] (bottom) polyads. The lines in the diagram represent
5 the state energies within the polyad in different limits. In the middle of the diagram are the eigenenergies, and at the far left and right are the zero-order energies of the normal mode and local mode basis sets, respectively. At positions intermediate between the eigenstate and basis state extremes, the energies are
10 calculated by diagonalizing the effective Hamiltonian with the off-diagonal elements multiplied by a scaling factor between 0 (the unperturbed basis set limit) and 1 (the eigenstate limit). An eigenstate is likely to be assignable in terms of quantum numbers associated with a given basis set if the line which passes from the
15 eigenstate to a zero-order basis state does so with minimal deviations, or can be followed through avoided crossings. It is clear from this diagram that many more eigenstates in the [22,0,*g*,e] polyad are assignable in the local mode basis set than in the traditional normal mode basis set, but that the normal mode
20 basis set provides a better zero-order description of the eigenstates in the [8,0,*g*,e] polyad. This is Fig. 8 from Jacobson et al.[21]

At the extreme left and right respectively are the normal
25 mode and local mode basis state energies. In the middle are the eigen-energies. The connecting lines are the energies obtained with *all of the off-diagonal matrix elements* of $\mathbf{H}^{eff}$ multiplied by a scaling factor that is varied from zero at the extremes to 1 at the centre. Connection
30 between the extreme and the centre without avoided crossings is a sign that the basis state quantum numbers remain good and that the $P_i$ and $D_j$ values remain near 1.

There are two spaghetti diagrams in Fig. 6, one for $N_{bend}$=8
35 and one for $N_{bend}$=22 (very close to the top of the acetylene-vinylidene barrier, which is located just above $N_{bend}$=24). At low $N_{bend}$, the normal mode basis set is best, especially at the top and bottom of the polyad. At high $N_{bend}$, the local mode basis set is a good basis, except for

40 the tangle of levels near the middle of the polyad. The top (counter-rotators) and bottom (local-benders) eigenstates are nearly pure basis states[38], with $P_i \approx 1$ and $D_j \approx 1$. A dramatic bifurcation occurs near $N_{bend}$=14. Pure-bending eigenstates emerge and become increasingly pure as $N_{bend}$
45 increases. These extreme local-benders have *gerade, ungerade* inversion symmetry, as is required for any molecule, such as HCCH, which has a centre of symmetry. For example, at $N_{bend}$=22, there is a *g,u* pair of eigenstates

50 $|N_{bend}=22,g\rangle = 2^{-1/2}[|v_{left}=22,v_{right}=0\rangle + |v_{left}=0,v_{right}=22\rangle]$
$|N_{bend}=22,u\rangle = 2^{-1/2}[|v_{left}=22,v_{right}=0\rangle - |v_{left}=0,v_{right}=22\rangle]$ (10).

The emergence of pure, extreme local-bender eigenstates is a clearly demonstrated property of an approximate,
55 reduced-dimension $\mathbf{H}^{eff}$ [39]. However, Hua Guo's high-level *ab initio* calculations of low-J *g↔u* vibrational transition strengths for the complete **H**, illustrated by Figs. 7 and 8 show that the emergence of local-benders is not an artefact that arises from the use of an approximate $\mathbf{H}^{eff}$.
60

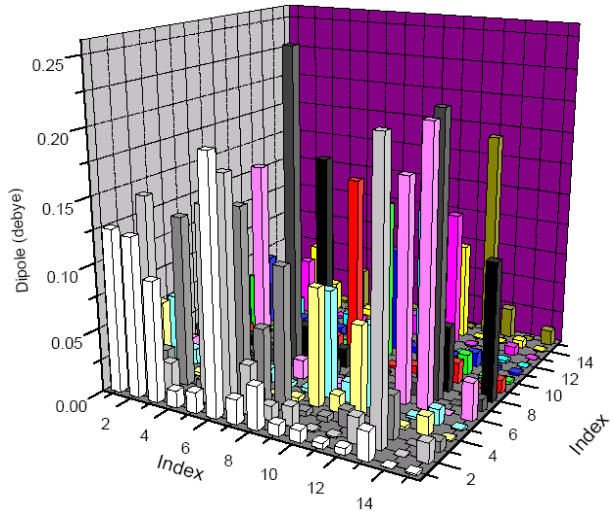

**Figure 7:** Acetylene $S_0$ *g-u* $\ell$=0-0 computed transition moments between low energy vibrational levels in the normal mode limit[39]. Notice that there are many weakly allowed transitions and most
65 transition moments are smaller than 0.1 Debye.

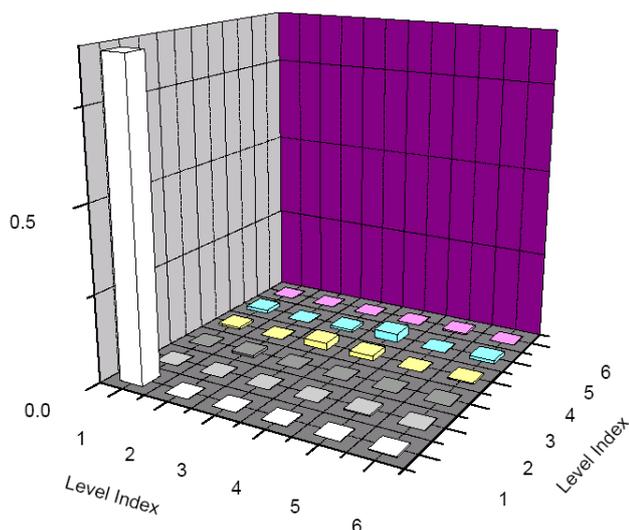

**Figure 8**: Computed vibrational transition moments between vibrational levels that are well into the local mode limit[39]. Notice that there is only one transition, [$N_{bend}=20, \ell_{total}=0, g$] to [$N_{bend}=20, \ell_{total}=0, u$], that has appreciable transition strength, ~1 Debye, and all of the other transition moments are <<0.1 Debye. Contrary to normal expectations of greater complexity at high vibrational energy, a dramatic simplification of the transition structure emerges.

This emergence of local-benders is part of a global reorganization of the structure and dynamics of $S_0$ acetylene. At low energy, vibrational transition probabilities are small and apparently indiscriminately distributed over many *g,u* pairs of rotation-vibration energy levels. At higher energy, all of the transition strength seems to coalesce into one strongly allowed $\Delta N_{bend}=0$, $g \leftrightarrow u$, $\Delta J=\pm 1$ pair of vibration-rotation transitions[39].

Owing to the near perfect degeneracy between the same-J, same-$N_{bend}$ pair of *g,u* local-bender vibrational levels[21-23], this *g-u* allowed vibration-rotation transition occurs at exactly the frequency of what would be a *g-g* forbidden pure-rotation transition in a local-bender state. Pure rotation transitions in $S_0$ acetylene are forbidden, but these local-bender vibration-rotation transitions are strongly allowed, with electric dipole transition moments in excess of 1 Debye.

The Field group designed its CPmmW spectrometer to observe a-dipole J=0-1 transitions in *both* $S_0$ local-bender acetylene and $S_0$ vinylidene, *both* expected to occur in the 70-85 GHz frequency region. We expected that 193 nm photolysis of vinyl cyanide would produce vinylidene, local-bender acetylene, HNC, and HCN, all of which have J=0-1 transitions within the tuning range of the spectrometer. We also expected that, even if vinylidene were produced at very high J, rotational cooling in the supersonic jet expansion would cool the rotations to $T_{rot} \sim 5K$. We were confident that the extreme purity of the local-bender levels of acetylene would ensure that redistribution of the local-bender character into the dense manifold of non-local-bender states would not occur. This confidence was based on the $\mathbf{H}^{eff}$ determined from fits to low-J levels[21-23].

The high-J $\mathbf{H}^{eff}$ fitted by Michel Herman's research group and subjected to dynamical analysis by David Perry and Michel Herman[24] reveals a dramatic destruction of the stability of extreme local-bender states at very high J. For example, the participation numbers for the extreme local-bender level in the $N_{bend}=16$ polyad are respectively 1.0034, 7.23, and 43.7 for J=2, 30, and 100. This turning on of Intramolecular Vibrational Redistribution at high-J is illustrated more globally in Figs. 9 and 10.

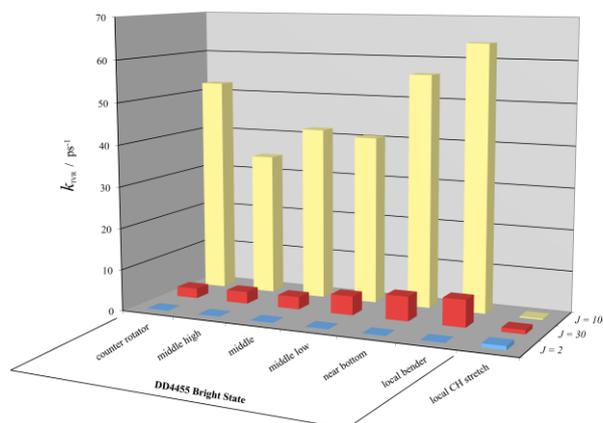

**Figure 9:** Initial rates of IVR in acetylene for bending bright states ($\ell_{total} = 0$, $N_{Stretch} = 0$) for various polyad members, expressed in the DD4455 basis set, and for the local CH stretch bright state. Data are for $J = 2$, 30 and 100 in the [$N_{bend}=16, g, e$] polyad. In the DD4455 basis, all of the vibrational interaction matrix elements at J = 0 have been prediagonalized; therefore, the plotted rates reflect only the rotationally mediated interactions, including the rotational *l*-type resonance and Coriolis interactions. This figure is Figure 10 from Perry et al.[24]. The rates are derived from a spectroscopic Hamiltonian fitted to 19,582 vibration-rotation transitions vibrational states with energies up to 13,000 cm$^{-1}$.

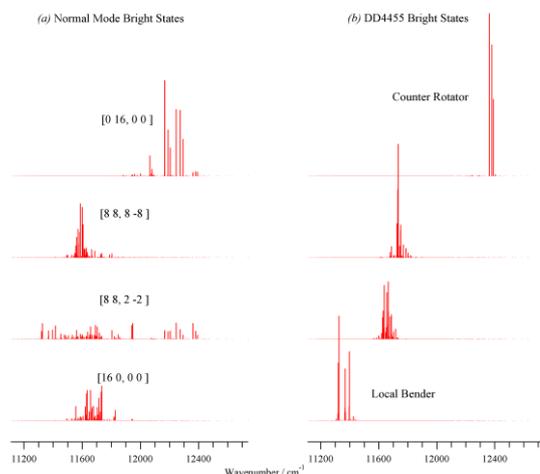

**Figure 10**: Simulated spectra of different bright states for $J = 30$ in the [16,*g,e*] polyad of acetylene. (a) Normal mode bright

states. (b) Bright states in the DD4455 basis set. The indicated wavenumbers include the vibrational energies plus about 1,120 cm$^{-1}$ of rotational energy. The fractionation in (a) results from the combination of anharmonic vibrational interactions plus the rotationally mediated interactions; whereas in (b) the fractionation arises entirely from the rotationally mediated interactions. This figure is Figure 8 from Perry et al.[24]

**F. CPmmW Spectroscopy of 193 nm Photolysis of Vinyl Cyanide**.

$C_2H_2$ ($S_0$ acetylene and vinylidene), $S_0$ HCN, and $S_0$ HNC are dominant products of 193 nm photolysis of vinyl cyanide. Pure rotation J=0-1 a-dipole spectra of local-bender vibrationally excited acetylene, vinylidene, HCN, and HNC are each in principle detectable in the Field group's CPmmW spectrometer. The J=0-1 transitions in many vibrational levels of each of these species are expected to be well resolved and, from the intensities of these transitions, we expected that it would be possible to determine a set of relative species-vibrational populations. However, in order to extract species-vibrational populations from our pure rotation CPmmW spectra, it is necessary to cool the nascent rotational distribution to a $T_{rot} \approx 5$ K *thermal* distribution in a supersonic expansion without significantly affecting the nascent species-vibrational population distribution. Our expectation, based on the low-J **H**$^{eff}$ [21] and the *ab initio* results that reveal the dramatic emergence of strong local-bender vibrational transitions[39], was that the three-centre reaction that forms HCN + vinylidene co-products populates low-lying vibrational levels of $S_0$ vinylidene. We expected that these vinylidene resonances would in-turn funnel population into nearby extreme local-bender vibrational levels of $S_0$ acetylene.

1. **Experimental details.** Supersonic cooling of the rotational degrees of freedom of photolysis products is essential. Since vibrational relaxation occurs on a much slower time-scale compared to that of rotational relaxation, it should be possible to control the amount of collisional cooling of photolysis products by precisely locating the photolysis laser beam within the collisional region of the adiabatic expansion. In an axial jet expansion from a 1 mm diameter nozzle, however, the number density decreases as the square of the distance from the nozzle, and the entire collisional region, in which rotational cooling occurs, is contained within a distance of only a few nozzle diameters from the nozzle orifice. In previous Field group experiments it was difficult to control the amount of rotational cooling by adjusting the location of the photolysis beam within the expansion region.

A slit jet, designed following drawings supplied by Terry Miller's group[40], is used in the experiments described here. Two General Valve solenoid drivers operate the poppet with an attached rubber cord, which seals the 5 cm long, 0.2 mm wide slit from inside of the slit jet body. Adiabatic expansion of the molecular beam produced by a slit jet is mostly one-dimensional with number density in the expanding gas decreasing linearly with the distance from the slit. The collisional region in our experiment extends 2–3 centimetres from the slit, is 3–4 cm wide, and can accommodate the unfocused photolysis laser beam with a cross section of $1 \times 2.5$ cm. Another advantage of the slit jet relative to the axial jet is a reduction of Doppler dephasing of the FID collected in a direction parallel to the slit. We observe a 150 kHz Doppler linewidth (in the 63-100 GHz region), which is a 2–3 fold improvement over what we could achieve with the axial jet. Clustering of the molecules expanded from a slit jet is more problematic than in an axial jet, thus greater dilution in carrier gas was required.

Vinyl cyanide (VCN) at $\geq$ 99% purity and its isotopologues at $\geq$ 98% purity are purchased from Sigma Aldrich and mixed, without further purification, with argon gas in a sample cylinder at 0.25% VCN in Ar. We find no difference in cyanovinyl radical signal strength when nitrogen is used instead of Ar as a carrier gas, but we obtain an order of magnitude weaker signal with helium, which we attribute to less efficient rotational cooling. The mixture is expanded from the slit jet, at a stagnation pressure of 150 mbar, into a vacuum chamber with approximately 60 L volume, which is pumped by a 2400 L/s turbo-molecular pump (Osaka TG2400MBWC), situated opposite to the slit jet. The turbo pump is backed by a 45 cfm two-stage rotary vane pump (Alcatel 2063). The pressure in the chamber, measured by a 0.05 Torr scale capacitance manometer (MKS 627DU5TDD1B), is $5 \times 10^{-4}$ mbar when the pulsed slit jet operates at a 20 Hz repetition rate. The gas pulse duration is estimated to be 700 μs when the on-time of the multi-channel IOTA ONE driver is set to 350 μs. A 193 nm ArF excimer laser (Lambda Physik COMPex 102) with 25–50 mJ/pulse energy photolyzes the VCN molecules 1.5 cm downstream from the slit. The mm-waves are loosely focused to an ~7 cm diameter beam by two 50 cm focal length Teflon lenses located inside the vacuum chamber with their optical axis 7.5 cm downstream from the slit. Two 2.5 cm thick Teflon windows, located adjacent to the lenses transmit the chirped mm-wave pulse into and the FID out of the vacuum system.

2. **Spectroscopic Results.** The 70-102 GHz operating region of our current CPmmW spectrometer is sufficient to capture *only* the J=0-1 transitions of *all two-heavy-atom* polar molecular products (except CN) of 193 nm photolysis of VCN. The J=1-2 transitions of all two-heavy-atom molecules occur at too high frequency for our spectrometer, thus there cannot presently be any direct measurement of $T_{rot}$ or even a determination of whether the rotational populations are thermalized. However, there are several different polar molecules (parent VCN, photofragments, and known impurities in our VCN sample) that have more than two heavy atoms. For these molecules, several consecutive J-J+1 transitions are observed. The rotational populations for these larger molecules appear to be thermalized and can be used to estimate $T_{rot}$ for the two-heavy-atom molecules. However, if a two-heavy-atom photofragment is formed at extremely high rotational excitation, there is no assurance that the intensity of a J=0-1 transition bears any relationship to a nascent vibrational population. In fact, rotational cooling might be too slow to transfer population into J=0 and 1 at a rate that is sufficient to yield a detectable J=0-1 transition at our operating conditions.

CPmmW spectra of HCN and HNC are shown in Fig. 12. The J=0-1 transitions in HCN and HNC, respectively, are labelled with downward-pointing blue and red arrows. The HCN vs. HNC line assignments are largely based on the presence of much larger eqQ quadrupolar hyperfine splittings in HCN than HNC. The magnitude of the eqQ constant is strongly dependent on the amount of excitation in the bending mode ($v_2$) and weakly dependent on excitation in the CN stretch mode ($v_3$)[7]. Vibrational assignments are based on the combination of the rotational constant, B, and eqQ. Globally, excitation of the bending mode increases B while excitation of the stretches decrease B. Some vibrational assignments are shown on Figs. 14 and 15, but with few exceptions, vibrational assignments are unambiguous. Two strong transitions near 88.48 GHz are assigned as belonging to the (0,2,1) and (1,2,0) vibrational levels[42]. The intensities of J=0-1 rotational transitions in these vibrational levels correspond to a much higher vibrational temperature than most of the bend overtone (0,$v_2$,0) levels sampled in our CPmmW spectra.

**Vinyl Cyanide Photolysis: 3 vs. 4 Centre Transition State**

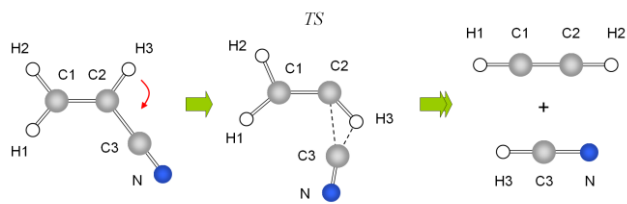

**Makes Vinylidene & Local-Bender Acetylene: strong mm-wave transitions ~75 GHz expected.**

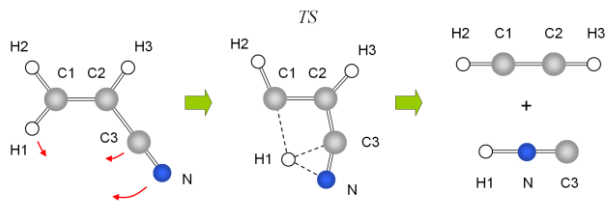

**Makes *cis*-bending Acetylene: no strong mm-wave transitions expected.**

**Figure 11:** A simple model for the relationships between the $C_2H_2$ and (HCN, HNC) photolysis co-products[27].

There is vastly more information about the HCN and HNC products of 193 nm photolysis of VCN than was available from the mass spectrometric plus photofragment translational spectroscopy results[6]. Figure 11 illustrates the simple model that had been used to organize the previous state-of-the-art experimental results[27]. The primary co-products of the three-centre transition state (1,1 elimination) are expected to be $S_0$ vinylidene, local-bend excited $S_0$ acetylene, and bend-excited HCN, while those of the four-centre transition state (1,2 elimination) are expected to be HNC and *cis*-bend excited acetylene. All three products of the three-centre mechanism are expected to be detectable in our CPmmW spectrometer. In contrast, only the HNC product of the four-centre mechanism is expected to be detectable. The *cis*-bend excited acetylene is expected to be anharmonically admixed into a dense manifold of essentially ergodic vibrational levels of acetylene, which will have undetectably weak J=0-1 transitions. This weakness is due to both small electric dipole transition moments and the distribution of population over too many vibrational levels[30].

Encouraged by the very strong HCN spectra, we searched very long and hard for the J=0-1 transitions in vinylidene and local-bender acetylene. At the bottom of Figure 13 is a spectrum that we initially believed was that of vinylidene, perhaps split into many components via interactions of an energetically remote acetylene local-bender doorway state with near degenerate ergodic acetylene vibrational levels.

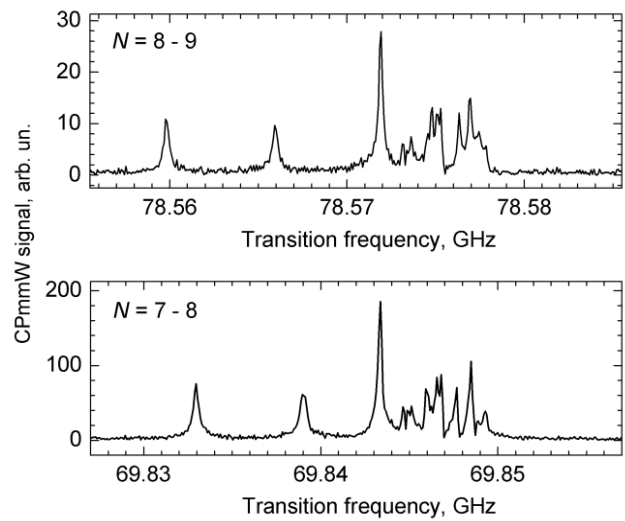

**Figure 13:** N=8 - 9 rotational transitions (top) and N=7 - 8 transitions (bottom) in the cyanovinyl radical.

How does one prove the molecular identity of a group of previously unobserved transitions in a mm-wave spectrum? High-level quantum chemical calculations of (B+C)/2= $\bar{B}$ rotational constants for vinylidene and local-bender acetylene[14] lead to predictions of J=0-1 transitions near 75 and 67 GHz, respectively. The observed cluster of lines near 70 GHz is unacceptably discrepant from the predictions. Use of $H_2{}^{13}C_2H^{13}CN$ $^{13}C$ isotopically substituted VCN results in a 2.55 GHz red-shift of the 69.84 GHz cluster that is compatible with vinylidene. The Zeeman effect for $S_0$ vinylidene should be very small, whereas that for a radical should be ~1 MHz/Gauss. Application of a few Gauss inhomogeneous magnetic field causes a few of the lines to vanish into baseline, but several of the lines appear to be resistant to shifting and broadening. This mixed verdict led us to the final test: [15]N substitution of the VCN. There is no N in vinylidene, yet the entire cluster of lines shifts 2.02 GHz to the red implying that the molecule contains one N atom. Utter failure! The final nail in the coffin is the rational multiple test. For a-dipole transitions in a near prolate top, J−(J+1) transitions occur at $(2\bar{B})(J+1)$. This means that if one transition is J−(J+1) the (J+1)−(J+2) transition will occur at a frequency larger by the factor (J+2)/(J+1). Figure 13 shows the near-perfect replication of the cluster of transitions at frequencies consistent with N=8 - 9 (top) and N=7 -8 (bottom).

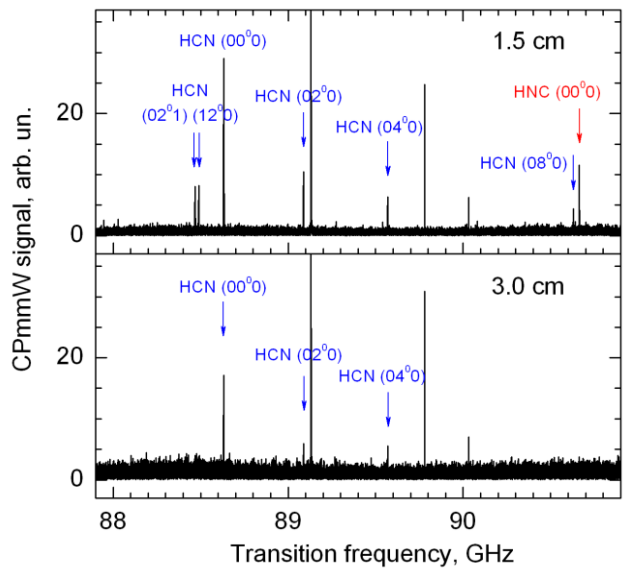

**Figure 14.** The effect of rotational cooling on the CPmmW spectrum of 193 nm photolysis products. When photolysis occurs 1.5 cm from the slit jet (top panel), significant rotational cooling occurs in the supersonic jet expansion. When photolysis occurs 3 cm downstream of the slit jet (bottom panel), rotational cooling is negligible. Notice the disappearance of several transitions in the 3 cm downstream spectrum owing to the failure of population to be relaxed all the way down to J=0 and 1.

Figure 14 shows the effects of a change in the amount of rotational cooling on the relative intensities of HCN and HNC J=0-1 transitions. When the 193 nm laser beam passes through the VCN beam at 1.5 cm downstream from the slit jet orifice, significant rotational cooling of the photofragments occurs. However, when the laser beam is shifted to 3 cm downstream, the expansion is nearly complete and very little rotational cooling occurs. It appears that reduced rotational cooling causes the J=0-1 transitions in the HCN stretch-bend combination vibrational levels and in all of the HNC vibrational levels to disappear. This suggests that the HCN stretch-bend combination levels and all of the HNC levels are formed rotationally much hotter than the HCN bending levels. Perhaps the degree of rotational excitation is a qualitative marker for fragmentation via distinct transition states?

Degree of rotational excitation is an important class of dynamical variable that is uniquely accessible to CPmmW spectroscopy. However, we have not had an opportunity to explore systematically variations in the extent of rotational cooling or to obtain spectra with S:N improved sufficiently to reveal the transitions that are unobservably weak in the presently obtained spectra. Systematic study of this rotational signature requires, at minimum, the ability to observe the relative intensities of J=0-1 *and* J=1-2 transitions.

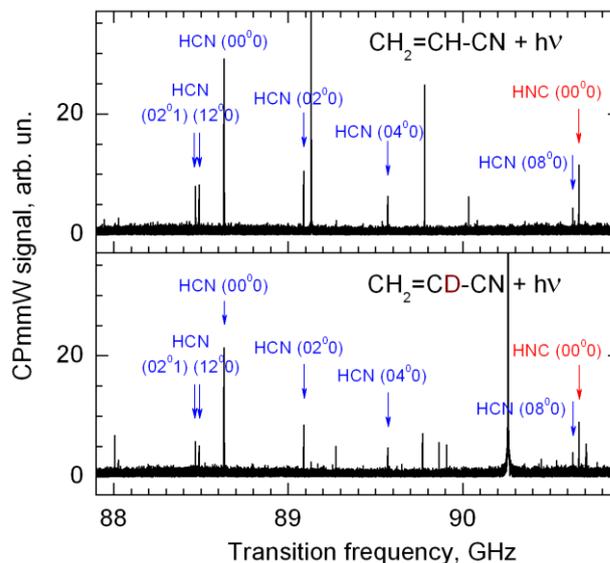

**Figure 15.** The effect of deuteration of the VCN parent molecule. Deuteration at the C-1 position should shut down the three-centre transition state for production of HCN. The small change in transition intensities upon deuteration implies that the simple three-centre picture is not valid.

Figure 15 shows the effect of partial deuteration of the VCN. If HCN is primarily formed via the 3-centre transition state (1,1 elimination) then H→D replacement on C-1 should significantly reduce the intensity of all HCN J=0-1 transitions, because DCN rather than HCN will be the primary photolysis product. DCN transitions (not shown) are detected, but at very different mm-wave frequencies. Figure 15 shows that photolysis of the D-substituted VCN parent molecule yields a reduction of only ~30% in the HCN transition intensities. It also appears that the HNC intensity (believed to be primarily a product of the four-centre transition state) is also reduced by ~30%. These are preliminary results, but they suggest that the three vs. four-centre picture is over-simplified[27,41]. Perhaps there is H-scrambling at the transition state? Perhaps there is CN roaming? The important point is that pure rotational spectroscopy introduces important new diagnostics for dynamics.

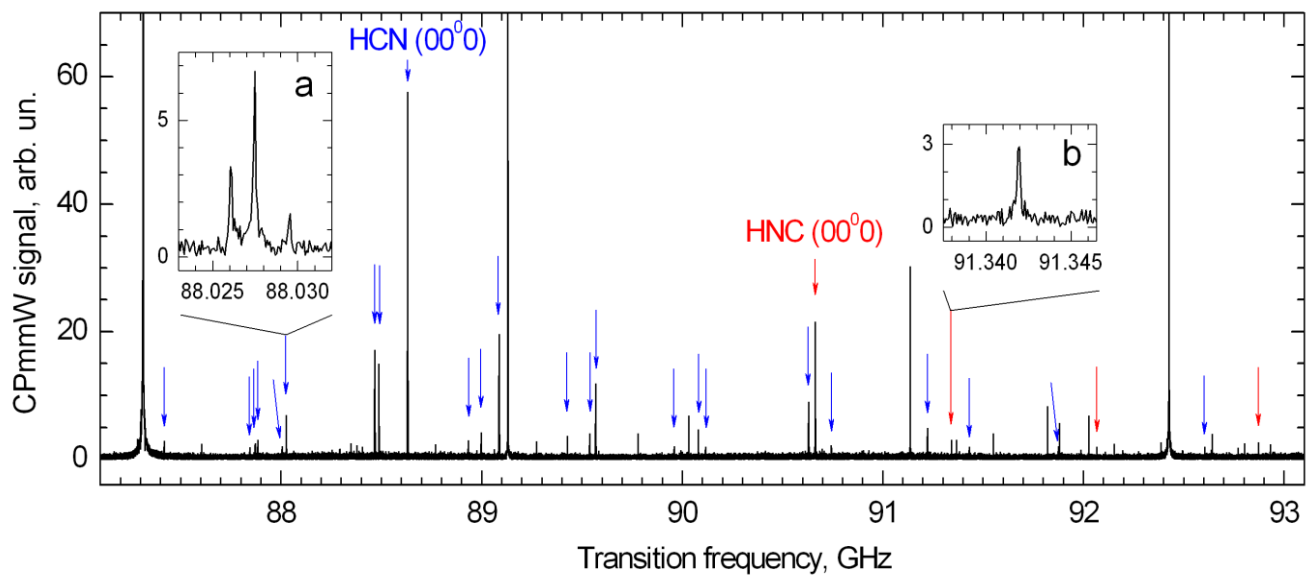

**Figure 12:** The CPmmW spectrum of products of 193 nm photolysis of VCN. The J=0-1 transitions in more than 25 HCN, HNC vibrational levels are identified. The insets show the characteristic eqQ hyperfine structure of HCN vs. HNC.

### G. Future Needs.

The essential weakness of pure rotational spectroscopy as a diagnostic for chemical dynamics is that, in order to provide a measure of nascent vibrational populations, it is necessary that rotation be thermalized. Although it seems reasonable to use $T_{rot}$, measured for parent or fragment "thermometer molecules" that contain more than two heavy atoms, as a measure of $T_{rot}$ for a two-heavy-atom photofragment, it is never possible to be sure that a molecule formed rotationally hot is rotationally cooled in the supersonic expansion sufficiently for the J=0-1 transition to be detectable in the CPmmW spectrum. At minimum, it is necessary to observe at least two rotational transitions in every populated vibrational level. Then the $T_{rot}$ inferred from the measured population ratio can be compared to the $T_{rot}$ of the thermometer molecule. If the two values of $T_{rot}$ are approximately equal, one can probably assume that rotational thermalization has occurred.

We failed to see the pure rotation spectrum of vinylidene. There are two likely reasons for this failure. One is that Franck-Condon restrictions invariably result in excitation of the parent molecule to energies well above the energy of the lowest photofragmentation transition state. Vibrationally mediated photodissociation[4] offers possibilities for excitation nearer threshold or for starting the initial wavepacket close to a selected conical intersection on the electronically excited potential energy surface. However, in the case of vinylidene, extreme rotational excitation is the enemy. We believe that, at low-J, vinylidene sees only a sparse manifold of extreme local-bender states of acetylene. At high-J, owing to strong Coriolis interactions between each extreme local-bender state and the dense manifold of "ergodic" acetylene states, the interaction between vinylidene and nearby acetylene states is promiscuous. It is Fermi Golden Rule statistical rather than selective. The relatively slow variation of vinylidene vibrational resonance widths in the PES[16] suggests that the interaction between vinylidene and vibrationally highly excited acetylene is not dependent on accidental near degeneracy of each vinylidene basis state with a local-bender basis state. The interaction is mediated by energetically remote local-bender doorway states[30]. Proof of this hypothesis will require an experimental scheme for photolysis of a Vinyl-X molecule that forms vinylidene at minimal initial rotational excitation. Infrared Multi-Photon Dissociation (IRMPD) using a $CO_2$ laser has been suggested[41]. This scheme could be optimized by vibrationally pre-heating the Vinyl-X in a pyrolysis nozzle or by selection of an X as a leaving group that is an exceptionally good absorber of $CO_2$ laser photons.

### H. Power of CPmmW in Photolysis that Produces Triatomic Co-Product

The CPmmW pure rotation spectra of HCN/HNC are extremely rich in information about the photolysis transition state(s). The vibrational and isomer specificity is unrivalled by other classes of chemical dynamics diagnostics. It is likely that all polar triatomic photofragments will carry previously unimaginable information about photolysis transition states, especially information about photolysis co-product(s).

## Conclusions

CPmmW spectroscopy captures a new class of chemical dynamics information about the 193 nm photolysis of Vinyl cyanide. It is clear that the photolysis mechanisms are more complicated than the simple 3-centre vs. 4-centre model depicted in Fig. 11[5,27,41]. Relative populations in many HCN, HNC vibrational levels are observed, but $S_0$ vinylidene has escaped detection. Rotation plays a critical role in the "stability" and high-resolution observability of vinylidene. When vinylidene is formed in low-J rotational levels, as in photodetachment, it is as stable as a highly excited vibrational level of a diatomic molecule. However, when vinylidene is formed rotationally highly excited, as in photolysis of V-X (X=Cl, Br, CN), it mixes into the acetylene vibrational quasi-continuum. Additional schemes for enhancing the capabilities of CPmmW spectroscopy and for observing the pure rotation spectrum of $S_0$ vinylidene are discussed.

## Acknowledgments

We are grateful to Dr. Gabriel Just and to Prof. T. A. Miller at The Ohio State University for sharing their expertise on the slit jet design, and to Dr. Michel Costes at the University of Bordeaux for providing us with the multi-channel IOTA ONE driver.

RWF thanks the Department of Energy (grant #DEFG0287ER13671) for primary support of this work (equipment and personnel support for KP, RGS, GBP, BMW), the Petroleum Research Fund (grant #50650-ND6) for support of KP, the Air Force Office of Scientific Research (contract #FA9550-09-1-0330) for support of MC, and the National Science Foundation (grant #1126380) for equipment and for partial support of GBP. We thank Joshua Baraban for many experimental and theoretical contributions.